\documentstyle[12pt,aaspp4,flushrt]{article}

\def\be{\begin{equation}}
\def\ee{\end{equation}}

\def\au{{\rm\,AU}}

\def\half{{\textstyle{1\over2}}}

\def\bfr{{\bf r}}

\def\omit#1{}
\def\pa{\partial}

\def\bfr{{\bf r}}

\def\bfR{{\bf R}}

\begin{document}

\title{Resonant capture by inward migrating planets}

\author{Qingjuan Yu and Scott Tremaine}

\affil{Princeton University Observatory, Peyton Hall, \\
Princeton, NJ~08544-1001, USA}

\begin{abstract}

\noindent
We investigate resonant capture of small bodies by planets that migrate
inwards, using analytic arguments and three-body integrations. If the orbits
of the planet and the small body are initially circular and coplanar, the
small body is captured when it crosses the 2:1 resonance with the planet. As
the orbit shrinks it becomes more eccentric, until by the time its semimajor
axis has shrunk by a factor of four, its eccentricity reaches nearly unity
($1-e\ll 10^{-4}$). In typical planetary systems, bodies in this
high-eccentricity phase are likely to be consumed by the central star. If they
can avoid this fate, as migration continues the inclination flips from 0 to
$i=180^\circ$; thereafter the eccentricity declines until the semimajor axis
is a factor of nine smaller than at capture, at which point the small body is
released from the 2:1 resonance on a nearly circular retrograde orbit. Small
bodies captured into resonance from initially inclined or eccentric orbits can
also be ejected from the system, or released from the resonance on highly
eccentric polar orbits ($i\simeq 90^\circ$) that are stabilized by a secular
resonance. We conclude that migration can drive much of the inner planetesimal
disk into the star, and that post-migration multi-planet systems may not be
coplanar.
\end{abstract}


\section{Introduction}

\noindent
Giant planets are found orbiting nearby stars with semimajor axes as small as
0.04\au\ (\cite{sch00}). The difficulty of forming planets so close to a star
has led to the suggestion that these planets formed at larger radii and
migrated inwards, perhaps through gravitational interactions with the gaseous
protoplanetary disk or planetesimal disk (\cite{lin96,mur98}). 

When massive planets migrate, they can capture smaller bodies in mean-motion
resonances.  Resonant capture in the solar system was invoked by
Goldreich (1965) \nocite{gol65}to explain the frequent occurrence of
commensurable periods among planetary satellites; in this case the satellites
are migrating outwards due to tidal friction (see \cite{pea99} for a
review). Similarly, the outward migration of Neptune probably led to the
capture of Pluto and many Kuiper-belt objects (``Plutinos'') into the 2:3
resonance with Neptune (Malhotra 1993, 1995\nocite{mal93,mal95}).

All known examples of resonant capture in the solar system involve outward
migration. However, the compact orbits of exosolar planets are believed to
result from inward migration. The purpose of this paper is to explore some of
the novel features of resonant capture during inward migration.

We shall use the Delaunay elements,
\be
L=(M_\star a)^{1/2}, \quad G=L(1-e^2)^{1/2}, \quad H=G\cos i, \quad \ell,
\quad g=\omega, \quad h=\Omega;
\label{eq:delaunay}
\ee
here $M_\star$ is the mass of the star in units where the gravitational
constant is unity, $a$, $e$, and $i$ are the semimajor axis, eccentricity, and
inclination; $\ell$, $\omega$ and $\Omega$ are the mean anomaly, argument of
periastron, and longitude of the ascending node; $G$ is the specific angular
momentum and $H$ is the $z$-component of the angular momentum.  In these
variables the Kepler Hamiltonian is
\be
H_K(L)=-{M_\star^2\over 2L^2},
\label{eq:kepham}
\ee
and the mean motion $n=\pa H_K/\pa L=M_\star^2/L^3=(M_\star/a^3)^{1/2}$. 

For simplicity we shall consider only the case of a test particle that is
captured by a massive body on a circular orbit with zero inclination. We shall 
call this body ``Jupiter''; its mass, semimajor axis, and mean motion are
denoted by $M_J$, $a_J$, and $n_J$. 

\section{Analytic results}

\label{sec:rescap}

\noindent
A $(p,q)$ mean-motion resonance occurs when $pn\simeq(p+q)n_J$, where $p>0$
and $q>-p$ are integers.  We distinguish inner resonances, which lie inside
Jupiter ($a<a_J$, $n>n_J$, $q>0$), from outer resonances ($a>a_J$, $n<n_J$,
$q<0$).

The Hamiltonian of the test particle can be written in the form
\be
H_K(L)+\sum_{p,q,m}A_{pqm}(L,G,H)\cos[p\ell+(p+q)(g+h-\phi_J)-2mg];
\label{eq:hamres}
\ee
here $\phi_J(t)=\int n_J(t)dt$ is the azimuth of Jupiter, and $p$, $q$, and
$m$ are integers. The angles $h$ and $\phi_J$ can only appear in the
combination $h-\phi_J$ because the Hamiltonian is invariant to changes in the
origin of the azimuth. The restriction that the coefficient of $g$ must be an
even integer $2m$ arises because the planetary potential is symmetric about
the equatorial plane, so that the Hamiltonian must be invariant under the
transformation $h\to h+\pi$, $g\to g+\pi$. For low-eccentricity,
low-inclination orbits, $|A_{pqm}|\propto (M_J/M_\star)e^{|q-2m|}i^{|2m|}$ for
$p,q\not=0$. The terms with $p+q=0$ can also contain a contribution from an
additional axisymmetric potential, due for example to a protoplanetary disk or
the equatorial bulge of the central star. Terms with $p=q=0$ are sometimes
called the secular Hamiltonian since they do not depend on either the mean
anomaly $\ell$ or the azimuth of Jupiter $\phi_J$ and thus are
time-independent for Keplerian motion. 

Each $(p,q)$ mean-motion resonance is represented in the Hamiltonian
(\ref{eq:hamres}) by an infinite series in $m$.  We can isolate the slowly
varying resonant angle by a canonical transformation to new momenta and
coordinates $(\bfR,\bfr)$ defined by the mixed-variable generating function
\be
S(\bfR,\ell,g,h)=R_1\ell + R_2g + R_3[p\ell + (p+q)(g+h-\phi_J)].
\label{eq:sssone}
\ee
We have
\be
L={\pa S\over\pa \ell}=R_1+pR_3,\quad G={\pa S\over\pa g}=R_2 + (p+q)R_3,\quad
H={\pa S\over\pa h}=(p+q)R_3,
\label{eq:rdef}
\ee
\be
r_1={\pa S\over \pa R_1}=\ell, \quad r_2={\pa S\over\pa R_2}=g, \quad r_3={\pa
S\over \pa R_3}=p\ell + (p+q)(g+h-\phi_J).
\label{eq:ssstwo}
\ee
Near the resonance, $r_1$ varies much more rapidly than $r_2$ or $r_3$, so we
can average the Hamiltonian over the fast variable $r_1$ at fixed $r_2$,
$r_3$.  The conjugate action $R_1$ is a constant of motion in the averaged
Hamiltonian and an adiabatic invariant of the original Hamiltonian.  Therefore
\be
C_{pq}\equiv(p+q)R_1=(p+q)L-pH=(M_\star a)^{1/2}[(p+q)-p(1-e^2)^{1/2}\cos i]
\label{eq:adinv}
\ee
is adiabatically invariant. 

The averaged Hamiltonian has two degrees of freedom. The nature of the
trajectories in the averaged Hamiltonian depends on the relative strength of
the secular and resonant Hamiltonians. If the secular Hamiltonian is either
large or small compared to the resonant Hamiltonian, the timescales for
libration or circulation of the associated coordinates $r_2$ and $r_3$ are
well-separated, and the problem can be reduced to one degree of freedom by a
second averaging. In the intermediate case, resonances with different $m$ can
overlap, leading to chaotic motion. All three regimes are present in various
solar-system contexts. 

Despite these complexities, some of the principal features of resonant capture
can be deduced directly from the adiabatic invariant $C_{pq}$
(eq. \ref{eq:adinv}). Suppose that the test particle is initially on a
circular, equatorial orbit at semimajor axis $a_i$. Then $C_{pq}=(M_\star
a_i)^{1/2}q$. Once the particle is captured into resonance, its semimajor axis
will migrate along with Jupiter's to preserve the mean-motion resonance,
thereby pumping up the test particle's eccentricity and/or inclination. The
adiabatic invariance of $C_{pq}$ then requires that
\be
\kappa\equiv(1-e^2)^{1/2}\cos i=1+{q\over p}\left[1-(a_i/a)^{1/2}\right].
\label{eq:kappadef}
\ee
The quantity on the left is initially unity (since $e=i=0$) and can only
decrease. For an outer resonance ($q<0$) this requires that $a_i/a<1$; in
other words capture into an outer resonance can only occur if the perturber
migrates outward. If the perturber migrates a long way outwards, then
$a_i/a\to 0$ and $\kappa\to
\kappa_\infty=1+q/p$. Note that $0<\kappa_\infty<1$ since $p>0$, $-p<q<0$;
thus captured orbits never achieve radial ($e=1$) or polar ($i=\half\pi$)
orbits.

Similarly, capture into an inner resonance ($q>0$) can only occur if the
perturber migrates inward. However, in this case the eccentricity and
inclination evolution is more dramatic. The quantity $\kappa\to 0$ when
$a=q^2a_i/(p+q)^2$; in other words, by the time Jupiter has migrated inward by
a factor of $q^2/(p+q)^2$ after capture, the test particle has been driven
onto either a radial or a polar orbit. If migration continues past this point
and the test particle remains in resonance, it must be driven onto a
retrograde orbit ($\kappa < 0$). The minimum possible value of $\kappa$ is
$-1$ (for $e=0$, $i=\pi$). This is attained when Jupiter has migrated inward
by a factor of $q^2/(2p+q)^2$ since capture; if migration continues past this
point the test particle cannot remain in resonance.

These simple considerations suggest that inward migration of a planet can
excite small planets or planetesimals into high-eccentricity,
high-inclination, or retrograde orbits. In the remainder of this paper we
shall investigate the evolution and fate of these objects numerically.

\section{Numerical integrations}

\noindent
In our integrations we use units in which the gravitational constant,
Jupiter's initial semimajor axis $a_{Ji}$ and orbital period are all unity, so
$M_\star=4\pi^2$.  Jupiter's mass is $M_J=0.001M_\star$.  Jupiter is assumed
to follow a circular orbit that migrates inward according to the rule
\be
a_J(t)=a_{Jf}+(1-a_{Jf})\exp(-t/\tau),
\ee
where $a_{Jf}=0.1$ and $\tau=1000$. The test-particle integrations were
carried out in regularized coordinates. 

\begin{figure}
\epsscale{0.8}
\plotone{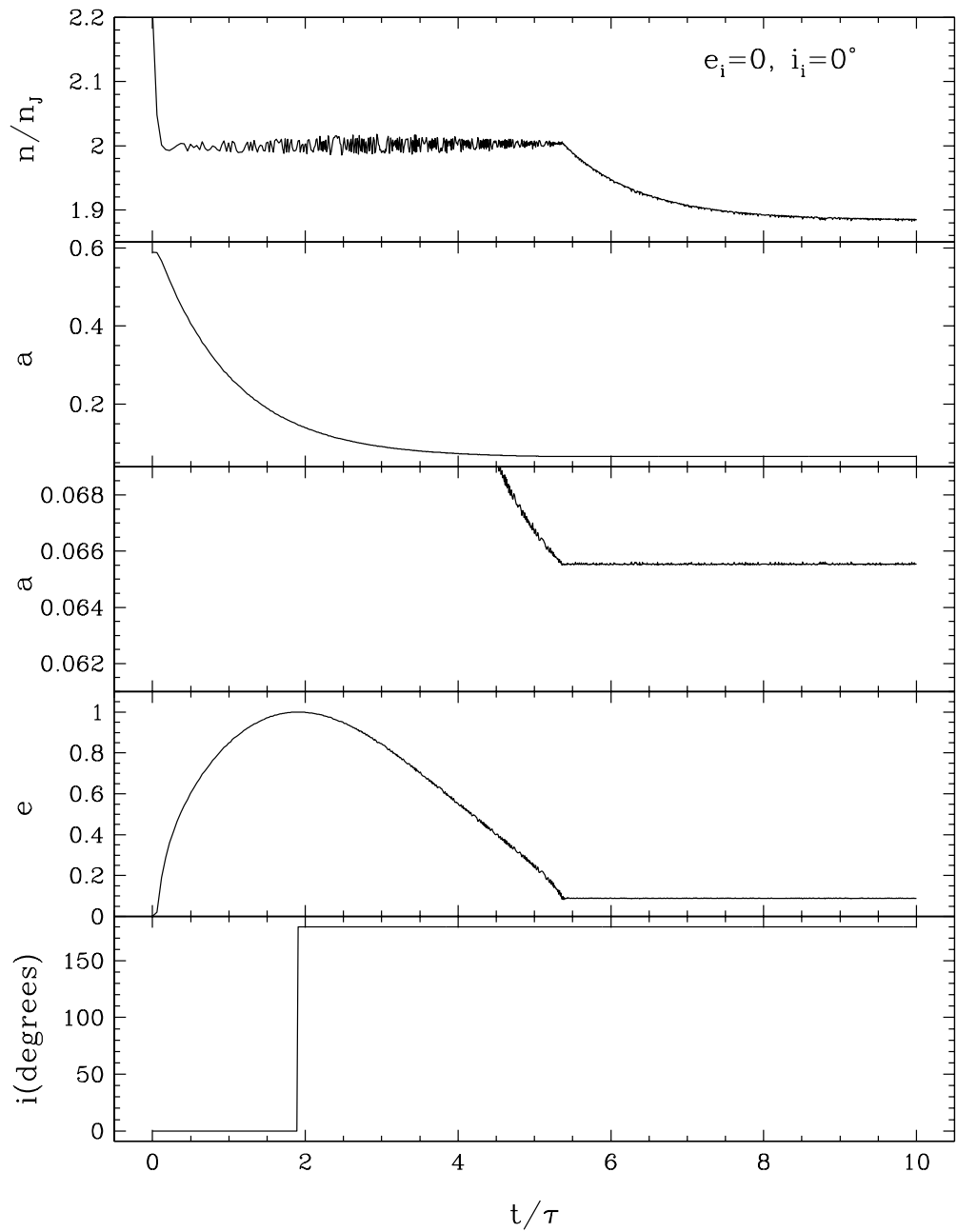}
\caption{
Resonant capture of a test particle by an inward migrating planet. From top to
bottom, the panels show the ratio of mean motions of Jupiter and the test
particle, the semimajor axis (two plots on different scales), the
eccentricity, and the inclination of the test particle. The test particle is
initially on a circular, equatorial orbit ($e_i=0$, $i_i=0$), shown as an
enlarged open circle in Figure \ref{fig:two}. It is captured into the 2:1
resonance with Jupiter and later released on a retrograde near-circular orbit
($i=180^\circ$, $e=0.09$).}
\label{fig:one}
\end{figure}

Figure \ref{fig:one} shows the evolution of a test particle initially on a
zero-inclination, circular orbit with semimajor axis $a_0=0.59$. The test
particle is captured into the 2:1 resonance ($p=q=1$) with Jupiter at
$t/\tau=0.073$. Thereafter its orbit shrinks and its eccentricity grows, as
predicted by the adiabatic invariant $\kappa$ (eq. \ref{eq:kappadef}).
Eventually at $t/\tau=1.90$, when the semimajor axis has shrunk by a factor of
$q^2/(p+q)^2=4$ since capture, the orbit is nearly radial
($1-e\ll10^{-4}$). In typical planetary systems, a small body in this
high-eccentricity phase would be vaporized by or collide with the star, but we
shall continue to follow the dynamical evolution assuming that the particle
survives.  At this point the inclination flips rapidly from $0$ to $\pi$
(recall that inclination is undefined for radial orbits). Once the inclination
is $\pi$ the eccentricity shrinks as the orbit continues to shrink, again as
predicted by adiabatic invariance. Finally, at $t/\tau=5.40$, when the
semimajor axis is a factor of $(1+2p/q)^2=9$ smaller than at capture, the test
particle is released from the resonance on a retrograde, equatorial,
near-circular orbit.

\begin{figure}
\epsscale{0.8}
\plotone{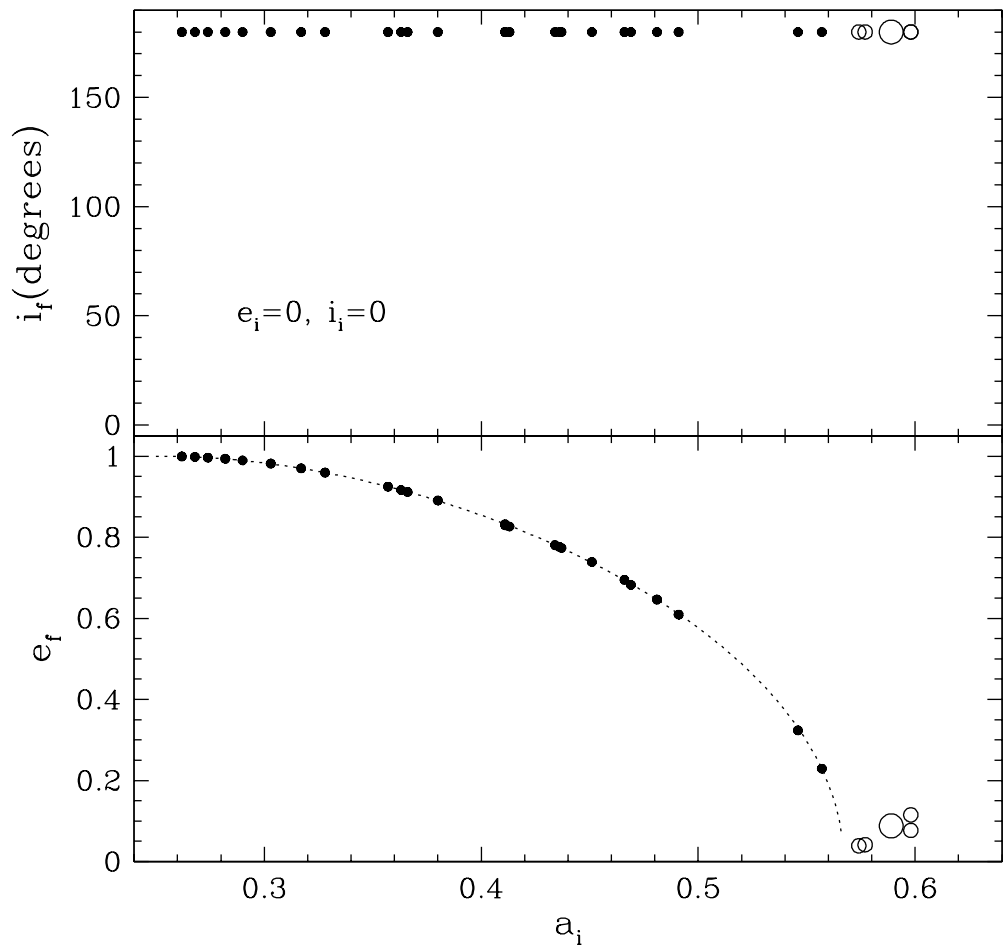}
\caption{
Resonant capture of test particles on initially circular, equatorial
orbits. The top and bottom panels show the final inclination and eccentricity,
$i_f$ and $e_f$, as a function of the initial semimajor axis $a_i$. Filled
circles represent test particles that are still in the 2:1 resonance at the
end of the integration, while open circles represent particles that have been
released from resonance. The dotted line shows the prediction of equation
(\ref{eq:predict}) for the final eccentricity. The enlarged open circle is the
test particle whose evolution is plotted in Figure \ref{fig:one}.}
\label{fig:two}
\end{figure}

Figure \ref{fig:two} shows the final states of an ensemble of test particles,
on initially circular, zero-inclination orbits with initial semimajor axes
$a_i$ between 0.26 and 0.6. The planet migrates from $a_{Ji}=1$ to
$a_{Jf}=0.1$, so the 2:1 resonance migrates from $a_{Ri}=0.6300$ to
$a_{Rf}=0.0630$. According to equation (\ref{eq:kappadef}), particles will be
captured into resonance during the migration if $a_i>a_{Rf}$, flipped onto
retrograde orbits if $a_i>(1+p/q)^2a_{Rf}=4a_{Rf}=0.2520$, and released from
resonance on nearly circular, retrograde orbits if
$a_i>(2p/q+1)^2a_{Rf}=9a_{Rf}=0.5670$.  For particles with
$4a_{Rf}<a_i<9a_{Rf}$ the eccentricity is given by
\be
a_i=a_{Rf}\left[2+(1-e_f^2)^{1/2}\right]^2,
\label{eq:predict}
\ee
which is plotted as a dotted line. All of these predictions agree very well
with the results shown in the Figure. 

\begin{figure}
\epsscale{0.6}
\plotone{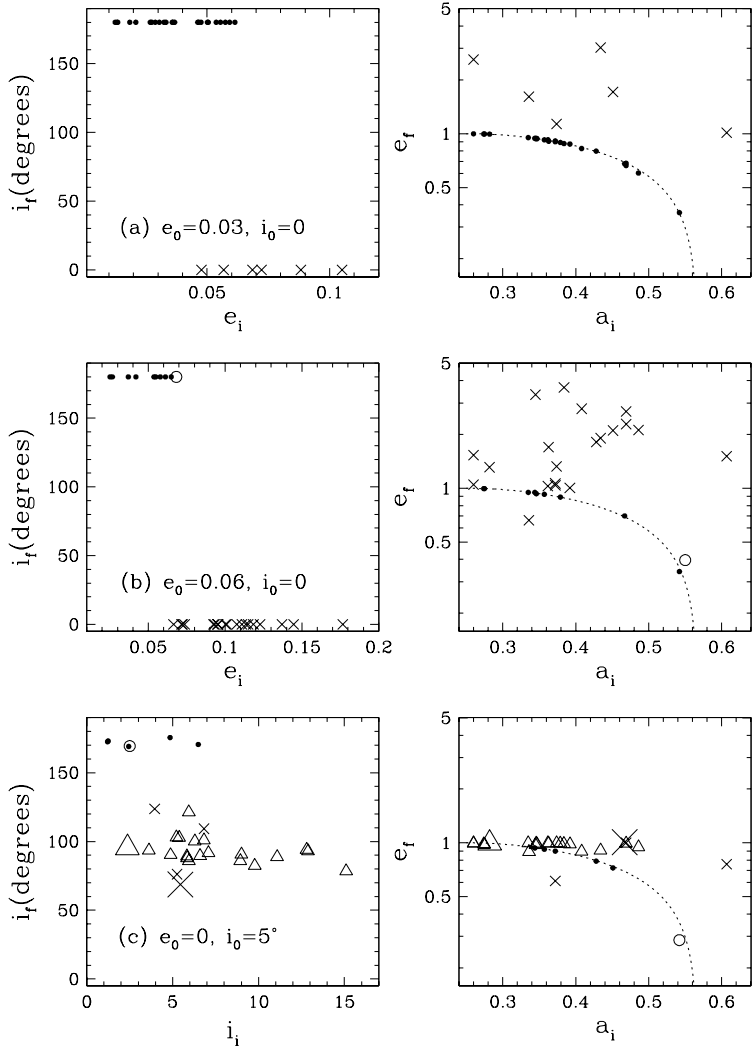}
\caption{
Evolution of test particles from low-eccentricity, low-inclination orbits. The
initial eccentricities and inclinations are chosen from the Schwarzschild
distribution (\ref{eq:ray}), with characteristic widths $e_0=0.03$, $i_0=0$
(top panels), $e_0=0.06$, $i_0=0$ (middle panels), and $e_0=0$, $i_0=5^\circ$
(bottom panels). The left- and right-hand columns show the final inclination
and eccentricity as a function of initial eccentricity or inclination and
initial semimajor axis. The dotted lines in the right-hand panels show the
prediction of equation (\ref{eq:predict}). The symbols denote the final states
of the test particles at the end of the integration: circles represent
retrograde orbits ($i\simeq \pi$); triangles represent nearly polar orbits
($i\simeq \half\pi$) and crosses represent escape orbits or chaotic orbits
that are likely to escape in the future. Filled symbols denote particles that
are in the 2:1 resonance at the end of the integration. Eccentricities and
inclinations are computed in the star-centered frame, except for escape
orbits, where the elements are computed in the center-of-mass frame. The
enlarged symbols in the bottom panels are the test particles whose evolution
is plotted in Figures \ref{fig:four} and
\ref{fig:five}. }
\label{fig:three}
\end{figure}

\begin{figure}
\epsscale{0.76}
\plotone{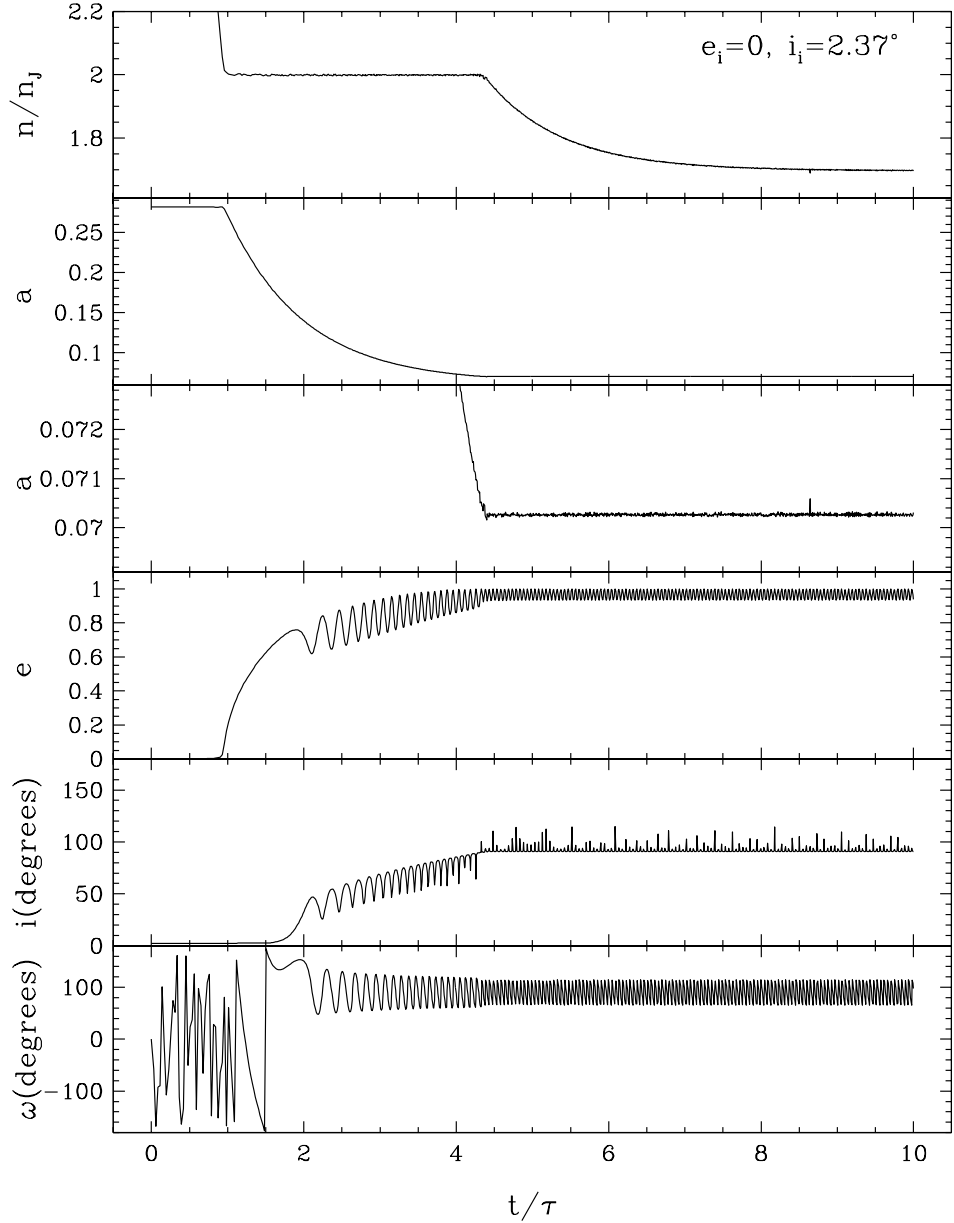}
\caption{
The evolution of a test particle with initial eccentricity and inclination
$e=0$, $i=2.37^\circ$, marked by an enlarged triangle in Figure
\ref{fig:three}. From top to bottom, the panels show the ratio of mean motions
of Jupiter and the test particle, the semimajor axis (two plots on different
scales), the eccentricity, the inclination, and the argument of periastron
(the angle in the orbital plane from the equatorial plane to the
periastron). The test particle is captured into the 2:1 resonance and later
released on a polar ($i\simeq \half\pi$) orbit. The particle is protected from
close encounters with the planet by a secular resonance, in which the argument
of periastron librates around $\half\pi$. }
\label{fig:five}
\end{figure}

The evolution is more complicated when the test particles have non-zero
eccentricities and inclinations. Figure \ref{fig:three} shows the final states
of test particles with the same range of initial semimajor axes as in Figure
\ref{fig:two}, but with eccentricities and inclinations chosen from the
Schwarzschild or Rayleigh distribution,
\be
n(e,i)de\,di\propto \exp\left(-{e^2\over 2e_0^2}-{i^2\over
2i_0^2}\right)e\,de\,i\,di. 
\label{eq:ray}
\ee
At the end of the integration, most of the particles are found in one of three 
final states:

\begin{itemize}

\item The particles shown as filled circles are in retrograde ($i\simeq\pi$)
orbits trapped in the 2:1 resonance, like those shown in Figure
\ref{fig:two}. If migration had continued, these would eventually be released
from resonance on nearly circular orbits, after their semimajor axis had
migrated by a factor 9 since capture (this has already happened for two
particles, marked by open circles). These particles typically 
have small initial eccentricities and inclinations, $e_i\lesssim 0.05$,
$i_i\lesssim5^\circ$. 

\item The particles marked by crosses are on escape orbits, or chaotic orbits
that are likely to escape in the future. In most cases
these particles are pumped onto high-eccentricity orbits, released from
resonance, and then suffer a close encounter with Jupiter, leading to a random
walk in semimajor axis and eventual escape. The evolution of the orbital
elements in a typical case is shown in Figure \ref{fig:four}. These particles
typically have initial eccentricities $e_i\gtrsim 0.05$. Note that the final
eccentricity and inclination for the escaping orbits are given in the
center-of-mass frame, while all other orbital elements in this and other
Figures are given in the star-centered frame.

\begin{figure}
\epsscale{0.75}
\plotone{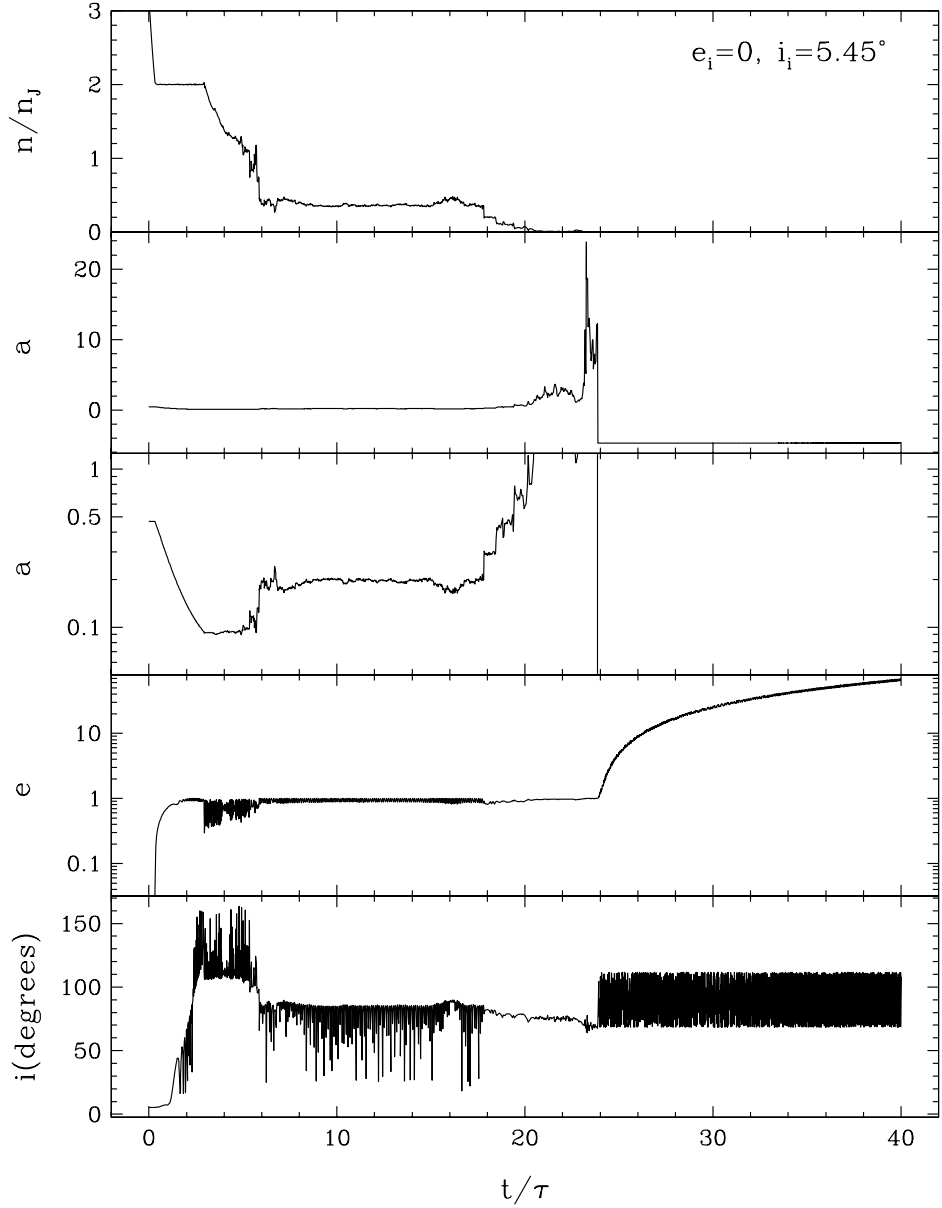}
\caption{
The evolution of a test particle with initial eccentricity and inclination
$e_i=0$, $i_i=5.45^\circ$, marked by an enlarged cross in Figure
\ref{fig:three}. From top to bottom, the panels show the ratio of mean motions
of Jupiter and the test particle, the semimajor axis (two plots on different
scales), the eccentricity, and the inclination of the test particle. The
particle is temporarily captured into the 2:1 resonance and later released on
a chaotic orbit, which eventually escapes after a close encounter with
Jupiter.  The eccentricity and inclination are not constant at large time
because we are working in a frame centered on the star rather than the center
of mass of the star and Jupiter. The final eccentricity and inclination in the
center-of-mass frame are $e=1.01$, $i=69^\circ$.}
\label{fig:four}
\end{figure}

\item The particles marked by open triangles were trapped in the 2:1
resonance and later released on nearly polar orbits ($i\simeq\half\pi$). These
particles have high eccentricities and arguments of perihelion $\omega$ that
librate around $\pm\half\pi$. This secular resonance protects
high-eccentricity orbits from close encounters with Jupiter, so the particle
orbits are stable on the timescale of our integrations ($1\times 10^4$ initial
orbital periods of Jupiter). These orbits typically have
initial inclination $i_i\gtrsim 5^\circ$. An example is shown in Figure
\ref{fig:five}.

\end{itemize}

\section{Discussion}

\noindent
Giant planets are found orbiting close to many nearby stars. Because the
conditions in the inner parts of protoplanetary disks are unfavorable to
planet formation, it is likely that these planets formed at much larger radii
and migrated inward. Moreover, because the characteristic evolution times in a
protoplanetary disk are longer at larger radius, it is likely that
planetesimals and small planets had already formed interior to this giant
planet before it began its inward journey. In these circumstances resonant
capture by the inward migrating planet is likely to occur at the 2:1
mean-motion resonance.

We have shown that small bodies that are captured from circular, equatorial
orbits into the 2:1 resonance will be excited onto nearly radial orbits by the
time their semimajor axis has shrunk by a factor of four, and released from
the resonance on nearly circular retrograde orbits when the semimajor axis has
shrunk by a factor of nine. In most of the known planetary systems, these
small bodies would be consumed by the central star during this
high-eccentricity phase. Thus, substantial migration of a giant planet is
likely to pollute the outer parts of the star with most of the mass in the
interior planetesimal disk: specifically, if the planet migrates from $a_{Ji}$ 
to $a_{Jf}<\frac{1}{4}a_{Ji}$, then the planetesimals with initial semimajor
axes in the range $2.520a_{Jf}<a<0.630a_{Ji}$ will be excited to
high-eccentricity orbits and are likely to be vaporized by or collide with the 
central star. 

Bodies that do not collide with the star will eventually be released from
resonance as migration continues. Some of these will be ejected, but some are
protected from ejection by secular resonance and can survive on polar,
high-eccentricity orbits. Others are released from resonance on circular and
equatorial but retrograde orbits, and will survive if the migration stops
before the giant planet reaches their semimajor axis. Thus, if the captured
planets survive the high-eccentricity phase, their post-migration
configuration may not be planar; if the orbits are coplanar, they may not all
orbit in the same direction.  If multiple bodies formed from a single disk
(``planets'') are not necessarily on coplanar orbits, they may be difficult to
distinguish from bodies formed from fragmentation of a collapsing gas cloud
(``stars'').  Radial-velocity observations cannot directly measure the
relative inclinations of planets in multi-planet systems; this determination
requires astrometric measurements such as those planned for the Space
Interferometry Mission (SIM).

This preliminary study has not addressed many important issues, including the
effects of an eccentric planet orbit, the stability of these resonances to
gravitational noise from other planets and massive planetesimals, or migration 
in systems containing two or more giant planets.

This research was supported in part by NASA Grant NAG5-7310.

\end{document}